\documentclass[titlepage,11pt]{article}
\usepackage{amsfonts}
\usepackage{amsmath}
\usepackage{amssymb}
\usepackage[onehalfspacing]{setspace}
\usepackage{geometry}
\usepackage{mathrsfs}
\usepackage{boxedminipage}

\newtheorem{theorem}{Theorem}

\newtheorem{corollary}[theorem]{Corollary}

\input{tcilatex}
\renewcommand{\mathcal}{\mathscr}

\begin{document}

\title{Trend and seasonality estimation for point-process time series}
\author{Daniel Gervini and Simon A.~Kopischke \\
Department of Mathematical Sciences\\
University of Wisconsin--Milwaukee\\
gervini@uwm.edu}
\maketitle

\begin{abstract}
This article introduces estimators of trend and seasonality for time series
of point processes. \ We assume the point processes follow a temporal or
spatial doubly-stochastic Poisson model with log-Gaussian intensity
functions. The proposed estimators are computationally simple M-estimators.
Their asymptotic distribution is derived, and their finite-sample
performance is studied by simulation. As an example of real-data
application, we study the patterns of bike demand in the Divvy bike-sharing
system of the city of Chicago.


\end{abstract}

\section{Introduction}

Point-process data is becoming increasingly common in statistical
applications. Although the use of point processes for data modeling is not
new (see e.g.~Cox and Isham, 1980; Snyder and Miller, 1991), the increasing
complexity and volume of data, as well as the more powerful computational
resources available today, allow for more sophisticated data modelling than
was possible in the past.

The literature on point processes has mostly focused on single realizations
of spatial, temporal or spatio-temporal processes (see e.g.~Diggle, 2013; M%
\o ller and Waagepetersen, 2004; Streit, 2010). Some of the examples
presented at the end of this paper could be analyzed that way. However, in
this paper we present an alternative, simpler approach, based on the ideas
of object-oriented data analysis (Marron and Dryden, 2021). In
object-oriented data analysis, one observes a collection of objects $\left\{
X_{t}:t=1,\ldots ,n\right\} $ which are something other than the usual
scalars or vectors of classical statistics; for example, they could be
continuous functions, as in functional data analysis (Ramsay and Silverman,
2005), or point processes, as in this paper. The advantage of this approach
is that the availability of $n$ replications of an object $X$ allows for
simpler statistical modelling and inference, more closely related to
standard univariate or multivariate statistical methods, than the somewhat
idiosyncratic techniques prevalent in the point-process literature.

Situations where a series of point processes $\left\{ X_{t}:t=1,\ldots
,n\right\} $ may arise include the slicing of a long temporal or
spatio-temporal process, say an annual process, into smaller time intervals,
such as daily time intervals. This way we obtain $n=365$ replications of a
point process. These daily replications, however, are unlikely to be
independent, so they should be seen as a time series, rather than
independent and identically distributed replications.

As an example of application, in this paper we will analyze the patterns of
bike demand in the Divvy bike-sharing system of the city of Chicago. Bike
sharing systems are common in large cities around the world (Shaheen et al.,
2010). They provide short-term bicycle rental services at unattended
stations distributed throughout the city. The Divvy system in Chicago keeps
records of every bike trip in the system, and makes the data publicly
available at the Chicago Data Portal (https://data.cityofchicago.org). The
set of bike check-out times for a given bike station can be seen as a
temporal point process. Slicing the annual set of check-out times into 365
daily sets allows us to better investigate and characterize the variability
of bike demand both within a day and across different days. For example, we
can answer questions like: is bike demand uniformly distributed during the
day, or does it peak at certain times? Is this pattern similar for every day
of the week, or is it different for weekends than for weekdays? Does bike
demand vary with the seasons, and if so, by what magnitude?

The above is an example of a time series of temporal processes. A time
series of spatial processes arises, for instance, when the times and
locations of certain crimes, like street theft, are recorded in a given city
on a given year (Ratcliffe, 2010), and then this spatiotemporal process is
sliced into 365 daily sets where, for each day, only the locations of the
incidents are kept.

The literature on statistical modelling of replicated point processes is
rather limited, and mostly deals with independent replications (Bouzas et
al., 2006, 2007; Fern\'{a}ndez-Alcal\'{a} et al., 2012; Wu et al., 2013;
Gervini, 2016, 2022a, 2022b; Gervini and Khanal, 2019; Gervini and Bauer,
2020; Xu et al., 2020a, 2020b; Gajardo and M\"{u}ller, 2022, 2023; Qiu et
al., 2024; Gajardo et al., 2025; Huang et al., 2025). However, as noted
above, replications arising from slicing will not be independent, so it is
necessary to develop new methods to address these dependencies. The existing
literature on log-Gaussian Cox processes is also extensive (Begu at al.,
2024; Diggle et al., 2005, 2013; M\o ller et al., 1998), however, it deals
with single realizations of processes. Although some of the examples
presented in this paper could be analyzed that way, the log-Gaussian models
presented here, which follow the object-oriented data analysis point of
view, are new.

Existing functional time series methods (Horv\'{a}th and Kokoszka, 2012,
ch.~16) cannot be directly applied to the type of data analyzed in this
paper. Functional data analysis assumes that the $X_{t}$s are smooth
functions, whereas in our context the $X_{t}$s are point processes. When the
number of observations per replication is high, the intensity functions
associated with the $X_{t}$s can be estimated by smoothing and then analyzed
using functional data methods, but this is not possible if the counts are
low. For point processes, the counts are intrinsically random and beyond the
researcher's control, so it is necessary to develop methods that are
specific for replicated point processes.

Recently, Gervini (2025) proposed a version of autocorrelation functions for
point-process time series. The data analyses in that paper show that the
presence of trends and seasonality are as common in point-process time
series as they are in traditional univariate or multivariate time series.
Unlike numerical time series, however, trends and seasonality cannot be
eliminated by differencing (Brockwell and Davis, 2006, ch.~1.4), since the
difference between two sets of points is not defined in any statistically
meaningful way.

The goal of this paper, then, is to propose methods for estimation of trends
and seasonal components for point-process time series. The processes can be
either temporal or spatial; the methods proposed here are based on splines,
which can be equally applied to temporal (univariate) or spatial (bivariate)
settings.

This paper is organized as follows. Section \ref{sec:Background} presents a
short background on point processes, in particular doubly-stochastic Poisson
processes. Section \ref{sec:Model} introduces additive log-Gaussian models
for trend and seasonality. Section \ref{sec:Estimation} derives parameter
estimators and Section \ref{sec:Asymptotics} develops the related asymptotic
theory. The finite-sample behavior of the estimators is analyzed by
simulation in Section \ref{sec:Simulations}, and a real-data application is
shown in Section \ref{sec:Examples}.

\section{Background on point processes\label{sec:Background}}

A point process $X$ is a random countable set in a space $\mathcal{S}$,
where $\mathcal{S}$ is usually $[0,+\infty )$ for temporal processes or $%
\mathbb{R}^{2}$ for spatial processes (M\o ller and Waagepetersen, 2004,
ch.~2). A point process is locally finite if, for any bounded set $%
B\subseteq \mathcal{S}$, $X\cap B$ is finite with probability one. In that
case, the count function $N(B)=\#(X\cap B)$ is well defined. Given $\lambda :%
\mathcal{S}\rightarrow \lbrack 0,\infty )$ such that $\int_{B}\lambda (u)du$
is finite for any bounded $B\subseteq \mathcal{S}$, the process $X$ is said
to be a Poisson process with intensity function $\lambda $, denoted by $%
X\sim \mathcal{P}(\lambda )$, if \emph{(i)} $N(B)$ follows a Poisson
distribution with rate $\int_{B}\lambda $ for any bounded $B\subseteq 
\mathcal{S}$, and \emph{(ii)} $N(B_{1}),\ldots ,N(B_{k})$ are independent
for any collection of disjoint bounded sets $B_{1},\ldots ,B_{k}$ in $%
\mathcal{S}$. A consequence of \emph{(i)} and \emph{(ii)} is that, for a
given bounded $B\subseteq \mathcal{S}$, the conditional distribution of the
points $X\cap B=\left\{ U_{1},\ldots ,U_{N(B)}\right\} $ given $N(B)=m$ is
the distribution of $m$ independent identically distributed random variables
with density $\lambda (u)/\int_{B}\lambda $.

From now on we will restrict our attention to realizations of the process on
a given bounded region $R\subseteq $ $\mathcal{S}$. Then, although in fact
we will be analyzing the truncated process $X\cap R$, for ease of notation
we will drop the $R$ and refer to $X\cap R$ simply as $X$. From the defining
properties \emph{(i)} and \emph{(ii)} of a Poisson process, then, it follows
that for $X\sim \mathcal{P}(\lambda )$ the density function of $X$ at a
realization $x=\{u_{1},\ldots ,u_{m}\}$ is given by 
\begin{equation}
f(x)=\exp \left( -\int_{R}\lambda \right) \frac{1}{m!}\prod_{j=1}^{m}\lambda
(u_{j}),  \label{eq:Pois_pdf}
\end{equation}%
where a density function, in this context, is understood as follows (M\o %
ller and Waagepetersen, 2004, Proposition 3.1): if $\mathcal{F}$ is the
family of finite subsets of $R$ and $h:\mathcal{F}\rightarrow \lbrack
0,\infty )$ is a set function, then 
\begin{equation}
E\{h(X)\}=\sum_{m=0}^{\infty }\int_{R}\cdots \int_{R}h(\{u_{1},\ldots
,u_{m}\})f(\{u_{1},\ldots ,u_{m}\})du_{1}\cdots du_{m}.  \label{eq:E_hX}
\end{equation}%
An example of set function is $h\left( x\right) =\sum_{j=1}^{m}g\left(
u_{j}\right) $ for a given $g:R\rightarrow \lbrack 0,\infty )$, with $%
h\left( \emptyset \right) =0$; in this case (\ref{eq:E_hX}) comes down to 
\begin{equation}
E\left\{ h\left( X\right) \right\} =\int_{R}g\left( u\right) \lambda \left(
u\right) du,  \label{eq:E_hg}
\end{equation}%
which is a particular case of Campbell's Theorem (Baddeley, 2007, ch.~2; M\o %
ller and Waagepetersen, 2004, ch.~4).

When a sequence of $n$ Poisson processes $\left\{ X_{t}:t=1,\ldots
,n\right\} $ is considered, a single intensity function $\lambda $ rarely
provides an adequate fit for all of them; it is more reasonable to assume a
specific intensity $\lambda _{t}$ for each $X_{t}$. This situation can be
best modelled by doubly-stochastic or Cox processes (M\o ller and
Waagepetersen, 2004, ch.~5), which are pairs $(X_{t},\Lambda _{t})$ where $%
\Lambda _{t}\left( u\right) $ is a continuous stochastic process (in the
variable $u$) that takes values in the space of nonnegative integrable
functions, and $X_{t}\mid (\Lambda _{t}=\lambda )\sim \mathcal{P}(\lambda )$%
. We will assume throughout this paper that the $X_{t}$s are conditionally
independent given the $\Lambda _{t}$s; therefore, the correlation structure
among the $X_{t}$s is entirely determined by the correlations among the $%
\Lambda _{t}$s. The realizations of the $X_{t}$s are observable but the $%
\Lambda _{t}$s are not, so they are treated as latent variables in the
models.

\section{Additive trend and seasonality models\label{sec:Model}}

We will assume that the latent processes $\{\Lambda _{t}:t=1,\ldots ,n\}$
are log-Gaussian, 
\begin{equation}
\log \Lambda _{t}\left( u\right) =\mu \left( u,t\right) +Y_{t}(u),
\label{eq:log-Gaus-model}
\end{equation}%
where $\left\{ Y_{t}(u):t=1,\ldots ,n\right\} $ are zero-mean Gaussian
processes on $R$, stationary in the variable $t$, and $\mu \left( u,t\right)
=E\left\{ \log \Lambda _{t}\left( u\right) \right\} $ is a function that, in
principle, may depend on both $u$ and $t$. By stationarity, the variance
function $v\left( u\right) =\mathrm{var}\left\{ Y_{t}\left( u\right)
\right\} $ and the autocovariance functions $\gamma _{k}\left( u,u^{\prime
}\right) =\mathrm{cov}\left\{ Y_{t}\left( u\right) ,Y_{t+k}\left( u^{\prime
}\right) \right\} $ do not depend on $t$. Then 
\begin{equation}
E\left\{ \Lambda _{t}\left( u\right) \right\} =\exp \left\{ \mu \left(
u,t\right) +\frac{1}{2}v\left( u\right) \right\}  \label{eq:ELmb}
\end{equation}%
and 
\begin{equation*}
E\left\{ \Lambda _{t}\left( u\right) \Lambda _{t+k}\left( u^{\prime }\right)
\right\} =E\left\{ \Lambda _{t}\left( u\right) \right\} E\left\{ \Lambda
_{t+k}\left( u^{\prime }\right) \right\} \exp \left\{ \gamma _{k}\left(
u,u^{\prime }\right) \right\} ,
\end{equation*}%
as shown in the Supplementary Material. It follows from (\ref{eq:E_hg}) and
the Law of Iterated Expectations that the counts $M_{t}=\#X_{t}$ satisfy $%
E\left( M_{t}\right) =\int_{R}E\left\{ \Lambda _{t}\left( u\right) \right\}
du$.

As mentioned in the Introduction, in many applications the counts $\left\{
M_{t}\right\} $ show a systematic trend over time; for example, for the
Divvy bike data, the overall bike demand is higher in summer than in winter.
Therefore, the function $\mu \left( u,t\right) $ in model (\ref%
{eq:log-Gaus-model}) does depend on $t$ in most applications.

The simplest way to model $\mu \left( u,t\right) $ is by an additive
decomposition, 
\begin{equation}
\mu \left( u,t\right) =\mu \left( u\right) +c\left( t\right) ,
\label{eq:trend_model}
\end{equation}%
where $c\left( t\right) $ is assumed to satisfy $c\left( 1\right) =0$ for
identifiability. Under model (\ref{eq:trend_model}), if we define 
\begin{equation}
\lambda \left( u\right) =\exp \left\{ \mu \left( u\right) +\frac{1}{2}%
v\left( u\right) \right\} ,  \label{eq:lmb}
\end{equation}%
it follows from (\ref{eq:ELmb}) that 
\begin{equation*}
E\left\{ \Lambda _{t}\left( u\right) \right\} =e^{c\left( t\right) }\lambda
\left( u\right) .
\end{equation*}%
We can interpret $\lambda \left( u\right) $ as the baseline intensity
function, since $\lambda \left( u\right) =E\left\{ \Lambda _{1}\left(
u\right) \right\} $, and then the time trend $c\left( t\right) $ introduces
a multiplicative factor $e^{c\left( t\right) }$ that increases or decreases
the overall magnitude of $E\left\{ \Lambda _{t}\left( u\right) \right\} $
with respect to the baseline $\lambda \left( u\right) $.

In addition to a time drift, the data sometimes show a natural periodicity;
for example, for daily processes, a 7-day periodicity is usually observed.
This can be accommodated by introducing additive seasonal components in the
model for $\mu \left( u,t\right) $. Let $d$ be the seasonal period. Then we
can write 
\begin{equation}
\mu \left( u,t\right) =\mu \left( u\right) +c\left( t\right) +s_{j(t)}\left(
u\right) ,  \label{eq:trend_plus_seas_model}
\end{equation}%
where $s_{1}\left( u\right) ,\ldots ,s_{d}\left( u\right) $ are the seasonal
components and $j\left( t\right) =\left\{ t\right\} _{d}$, where $\left\{
t\right\} _{d}$ is defined as the residue of the integer division of $t$ by $%
d$ if the residue is positive and $\left\{ t\right\} _{d}=d$ if $t$ is a
multiple of $d$. For identifiability, we assume $\sum_{j=1}^{d}s_{j}\left(
u\right) =0$. As before, we can best interpret model (\ref%
{eq:trend_plus_seas_model}) in terms of $E\left\{ \Lambda _{t}\left(
u\right) \right\} $. If $\phi _{j}\left( u\right) =\exp \left\{ s_{j}\left(
u\right) \right\} $, we have 
\begin{equation*}
E\left\{ \Lambda _{t}\left( u\right) \right\} =e^{c\left( t\right) }\phi
_{j}\left( u\right) \lambda \left( u\right) ,
\end{equation*}%
where $\lambda $ is as in (\ref{eq:lmb}). Therefore, the seasonal components
introduce a deformation on the shape of $E\left\{ \Lambda _{t}\left(
u\right) \right\} $ compared to the baseline $\lambda \left( u\right) $, in
addition to the change in overall magnitude introduced by $e^{c\left(
t\right) }$.

For estimation purposes it is more practical to reparameterize model (\ref%
{eq:trend_plus_seas_model}) using 
\begin{equation}
\mu _{j}\left( u\right) =\mu \left( u\right) +s_{j}\left( u\right) ,\ \
j=1,\ldots ,d,  \label{eq:mu_j}
\end{equation}%
since these $\mu _{j}(u)$'s are unconstrained, we recover $\mu \left(
u\right) $ and the $s_{j}\left( u\right) $'s from the $\mu _{j}(u)$'s as $%
\mu \left( u\right) =\sum_{j=1}^{d}\mu _{j}\left( u\right) /d$ and $%
s_{j}\left( u\right) =\mu _{j}\left( u\right) -\mu \left( u\right) $. Using
this reparameterization we have 
\begin{equation*}
E\left\{ \Lambda _{t}\left( u\right) \right\} =e^{c\left( t\right) }\lambda
_{j(t)}\left( u\right) ,
\end{equation*}%
where 
\begin{equation*}
\lambda _{j}\left( u\right) =\exp \left\{ \mu _{j}\left( u\right) +\frac{1}{2%
}v\left( u\right) \right\} ,~~j=1,\ldots ,d,
\end{equation*}%
are the seasonal baseline intensities.

For long time series, the trend $c\left( t\right) $ in model (\ref%
{eq:trend_plus_seas_model}) will usually show some periodicity as well,
although on a larger time span than the seasonality. For example, for the
daily Divvy data, the periodicity of $c\left( t\right) $ is annual, since it
is essentially a function of temperature, while the seasonal periodicity is
weekly. For this reason, to facilitate estimation and the asymptotic
analysis, we will assume that $c\left( t\right) =\tilde{c}(\left\{ t\right\}
_{r})$ for some integer period $r$ and some function $\tilde{c}\left(
t\right) $ defined on the interval $\left[ 1,r\right] $.

\section{Estimation of trend and seasonality\label{sec:Estimation}}

The functions $\mu \left( u\right) $, $s_{j}\left( u\right) $'s and $c\left(
t\right) $ in models (\ref{eq:trend_model}) and (\ref%
{eq:trend_plus_seas_model}) are estimated from the data. To do this in a
practical and flexible way, we use piecewise polynomials (splines) for $\mu
(u)$ and the $s_{j}\left( u\right) $'s. We assume, then, that $\mu \left(
u\right) =\mathbf{\theta }^{T}\mathbf{\beta }\left( u\right) $ and $%
s_{j}\left( u\right) =\mathbf{\xi }_{j}^{T}\mathbf{\beta }\left( u\right) $,
where $\mathbf{\beta }\left( u\right) $ is a vector of spline basis
functions on $R$ (for example, B-splines for temporal processes or
tensor-product B-splines for spatial processes) and $\mathbf{\theta }$\ and
the $\mathbf{\xi }_{j}$s are the spline coefficients. The identifiability
constraint on the seasonal components implies that $\sum_{j=1}^{d}\mathbf{%
\xi }_{j}=\mathbf{0}$. Using the alternative parameterization (\ref{eq:mu_j}%
), we have $\mu _{j}\left( u\right) =\mathbf{\theta }_{j}^{T}\mathbf{\beta }%
\left( u\right) $, with unconstrained $\mathbf{\theta }_{j}$s.

The trend $c(t)$, in general, tends to be a simple function that can be
adequately modeled by global polynomials. Therefore we let $c\left( t\right)
=\tilde{c}(\left\{ t\right\} _{r})$ for $\tilde{c}\left( t\right) $ a global
polynomial on $\left[ 1,r\right] $, which we can write as $\tilde{c}\left(
t\right) =\mathbf{\eta }^{T}\mathbf{b}(t)$ where $\mathbf{b}\left( t\right) $
is the basis of monomials centered at 1 (so that $\mathbf{b}(1)=\mathbf{0}$,
guaranteeing the identifiability constraint $\tilde{c}\left( 1\right) =0$
for any $\mathbf{\eta }$). This notation, as well as the theory in Section %
\ref{sec:Asymptotics}, allow for more general bases $\mathbf{b}\left(
t\right) $ than global polynomial, such as B-splines, but we will not use
them in this paper.

To estimate the parameters $(\mathbf{\theta },\mathbf{\eta })$ for model (%
\ref{eq:trend_model}) or $\left( \mathbf{\theta }_{1},\ldots ,\mathbf{\theta 
}_{d},\mathbf{\eta }\right) $ for model (\ref{eq:trend_plus_seas_model}) one
would normally use maximum likelihood estimation, but that would require
assumptions about the autocovariance functions of the $Y_{t}$s in model (\ref%
{eq:log-Gaus-model}) that may be hard to justify. Therefore, we will use a
working-likelihood approach, or M-estimation approach (Huber and Ronchetti,
2009, ch.~3): the estimators will be the maximum likelihood estimators for
model (\ref{eq:log-Gaus-model}) when $Y_{t}\left( u\right) =0$, although we
do not actually assume that the data follows such model. Then $(\mathbf{\hat{%
\theta}},\mathbf{\hat{\eta}})$ for the trend-only model (\ref{eq:trend_model}%
) is the maximizer of 
\begin{equation}
\rho _{n}\left( \mathbf{\theta },\mathbf{\eta }\right) =
\label{eq:rho_n_trend}
\end{equation}%
\begin{equation*}
-\frac{1}{n}\sum_{t=1}^{n}e^{\mathbf{\eta }^{T}\mathbf{b}(\left\{ t\right\}
_{r})}\int_{R}e^{\mathbf{\theta }^{T}\mathbf{\beta }\left( u\right) }du+%
\frac{1}{n}\sum_{t=1}^{n}m_{t}\mathbf{\eta }^{T}\mathbf{b}(\left\{ t\right\}
_{r})+\frac{1}{n}\sum_{t=1}^{n}\sum_{j=1}^{m_{t}}\mathbf{\theta }^{T}\mathbf{%
\beta }\left( u_{tj}\right) ,
\end{equation*}%
and $(\mathbf{\hat{\theta}}_{1},\ldots ,\mathbf{\hat{\theta}}_{d},\mathbf{%
\hat{\eta}})$ for the trend-plus-seasonality model (\ref%
{eq:trend_plus_seas_model}) is the maximizer of 
\begin{equation}
\rho _{n}\left( \mathbf{\theta }_{1},\ldots ,\mathbf{\theta }_{d},\mathbf{%
\eta }\right) =  \label{eq:rho_n_trend_seas}
\end{equation}%
\begin{equation*}
-\frac{1}{n}\sum_{t=1}^{n}e^{\mathbf{\eta }^{T}\mathbf{b}(\{t\}_{r})}%
\int_{R}e^{\mathbf{\theta }_{j(t)}^{T}\mathbf{\beta }\left( u\right) }du+%
\frac{1}{n}\sum_{t=1}^{n}m_{t}\mathbf{\eta }^{T}\mathbf{b}(\{t\}_{r})+\frac{1%
}{n}\sum_{t=1}^{n}\sum_{k=1}^{m_{t}}\mathbf{\theta }_{j(t)}^{T}\mathbf{\beta 
}(u_{tk}).
\end{equation*}%
These functions are concave, as shown in the Supplementary Material, so
their maximizations are computationally simple problems.

\section{Asymptotics\label{sec:Asymptotics}}

In this section we will establish the asymptotic consistency and Normal
distribution of the estimators when the number of replications $n$ goes to
infinity. We will first establish the asymptotics for the
trend-plus-seasonality model (\ref{eq:trend_plus_seas_model}) and then
derive the asymptotics for the trend-only model (\ref{eq:trend_model}) as a
corollary.

We will make the following assumptions:

\begin{description}
\item[A1] $X_{t}\mid (\Lambda _{t}=\lambda _{t})\sim \mathcal{P}(\lambda
_{t})$ and the $X_{t}$s are conditionally independent given the $\Lambda
_{t} $s, for $t=1,\ldots n$.

\item[A2] The $\Lambda _{t}$s follow model (\ref{eq:log-Gaus-model}) where
the $Y_{t}$s are zero-mean Gaussian processes, stationary in the variable $t$%
, with autocovariance functions $\gamma _{k}\left( u,u^{\prime }\right) $, $%
k\in \mathbb{Z}$.

\item[A3] The mean function $\mu \left( u,t\right) $ follows model (\ref%
{eq:trend_plus_seas_model}) with $c_{0}\left( t\right) =\tilde{c}_{0}\left(
\left\{ t\right\} _{r}\right) $ and $\mu _{0j}\left( u\right) =\mu
_{0}\left( u\right) +s_{0j}\left( u\right) $, where $\tilde{c}_{0}\left(
t\right) =\mathbf{\eta }_{0}^{T}\mathbf{b}(t)$ for $\mathbf{\eta }_{0}\in 
\mathbb{R}^{q}$ and $\mu _{0j}\left( u\right) =\mathbf{\theta }_{0j}^{T}%
\mathbf{\beta }\left( u\right) $ for $\mathbf{\theta }_{0j}\in \mathbb{R}%
^{p} $, $j=1,\ldots ,d$ ($\mathbf{\beta }\left( u\right) $ and $\mathbf{b}(t)
$ were defined in Section \ref{sec:Estimation}.) Moreover, we assume that $%
q<r$ and that $r$ is a multiple of $d$.

\item[A4] The variance function $v_{0}\left( u\right) =\mathrm{var}\left\{
Y_{t}\left( u\right) \right\} $ satisfies $v_{0}\left( u\right) =\mathbf{%
\tau }_{0}^{T}\mathbf{\beta }\left( u\right) $ for $\mathbf{\tau }_{0}\in 
\mathbb{R}^{p}$.

\item[A5] The autocovariance functions satisfy $\lim_{k\rightarrow +\infty
}\left\Vert \gamma _{k}\right\Vert _{\infty }=0$, where $\left\Vert \cdot
\right\Vert _{\infty }$ is the $\sup $ norm over $R\times R$.

\item[A6] $\gamma _{k}\left( u,u^{\prime }\right) =0$ for $k\neq 0$.
\end{description}

Let $(\mathbf{\hat{\theta}}_{1},\ldots ,\mathbf{\hat{\theta}}_{d},\mathbf{%
\hat{\eta}})$ be the estimators defined as the minimizers of the function $%
\rho _{n}\left( \mathbf{\theta }_{1},\ldots ,\mathbf{\theta }_{d},\mathbf{%
\eta }\right) $ in (\ref{eq:rho_n_trend_seas}). Since we assume $r=wd$ for
some integer $w$, we can write $\left\{ t\right\} _{r}$ as $a_{ij}=\left(
i-1\right) d+j$ for $i=1,\ldots ,w$ and $j=1,\ldots ,d$. Then, as shown in
the Supplementary Material, the function $\rho _{n}\left( \mathbf{\theta }%
_{1},\ldots ,\mathbf{\theta }_{d},\mathbf{\eta }\right) $ converges
pointwise in probability to 
\begin{equation}
\rho _{0}\left( \mathbf{\theta }_{1},\ldots ,\mathbf{\theta }_{d},\mathbf{%
\eta }\right) =  \label{eq:rho0_trend_seas}
\end{equation}%
\begin{equation*}
-\frac{1}{r}\sum_{i=1}^{w}\sum_{j=1}^{d}e^{\mathbf{\eta }^{T}\mathbf{b}%
(a_{ij})}\int_{R}e^{\mathbf{\theta }_{j}^{T}\mathbf{\beta }\left( u\right)
}du+\frac{1}{r}\sum_{i=1}^{w}\sum_{j=1}^{d}e^{\tilde{c}_{0}\left(
a_{ij}\right) }e_{j}\{\mathbf{\eta }^{T}\mathbf{b}(a_{ij})+\mathbf{\theta }%
_{j}^{T}\mathbf{\sigma }_{j}\},
\end{equation*}%
where 
\begin{equation}
\mathbf{\sigma }_{j}=\int_{R}\lambda _{0j}\left( u\right) \mathbf{\beta }%
(u)du  \label{eq:sigma_j}
\end{equation}%
and 
\begin{equation}
e_{j}=\int_{R}\lambda _{0j}\left( u\right) du  \label{eq:e_j}
\end{equation}%
with 
\begin{equation}
\lambda _{0j}\left( u\right) =\exp \left\{ \mu _{0j}\left( u\right) +\frac{%
v_{0}\left( u\right) }{2}\right\} .  \label{eq:lmb_0j}
\end{equation}%
Function (\ref{eq:rho0_trend_seas}) is strictly concave and maximized by $%
\left( \mathbf{\theta }_{01}^{\ast },\ldots ,\mathbf{\theta }_{0d}^{\ast },%
\mathbf{\eta }_{0}\right) $ defined in the next theorem, which does not make
use of assumption A6. Let $\hat{\lambda}_{j}\left( u\right) =\exp \{\mathbf{%
\hat{\theta}}_{j}^{T}\mathbf{\beta }\left( u\right) \}$ and $\widehat{\tilde{%
c}}(t)=\mathbf{\hat{\eta}}^{T}\mathbf{b}\left( t\right) $.

\begin{theorem}
\label{thm:Consistency_trend_plus_seas}Let $\mathbf{\theta }_{0j}^{\ast }=%
\mathbf{\theta }_{0j}+\mathbf{\tau }_{0}/2$ for $j=1,\ldots ,d$. Under
assumptions A1--A5 we have $\mathbf{\hat{\eta}}\overset{P}{\longrightarrow }%
\mathbf{\eta }_{0}$ and $\mathbf{\hat{\theta}}_{j}\overset{P}{%
\longrightarrow }\mathbf{\theta }_{0j}^{\ast }$ as $n\rightarrow \infty $,
for $j=1,\ldots ,d$. Therefore $\widehat{\tilde{c}}(t)\overset{P}{%
\longrightarrow }\tilde{c}_{0}\left( t\right) $ and $\hat{\lambda}_{j}\left(
u\right) \overset{P}{\longrightarrow }\lambda _{0j}\left( u\right) $
pointwise as $n\rightarrow \infty $, for $j=1,\ldots ,d$.
\end{theorem}

For the asymptotic normality, we assume that the $Y_{t}$s are independent.
However, this assumption is made just to simplify the asymptotic variance
formulas. Essentially the same proof given in the Supplementary Material can
be adapted to establish asymptotic normality under assumption A5 only, using
the Martingale Central Limit theorem (Pollard 1984, ch. VIII).

\begin{theorem}
\label{thm:AN_trend_plus_seas}Under assumptions A1--A6, we have 
\begin{equation*}
\sqrt{n}(\mathbf{\hat{\theta}}_{1}-\mathbf{\theta }_{01}^{\ast },\ldots ,%
\mathbf{\hat{\theta}}_{d}-\mathbf{\theta }_{0d}^{\ast },\mathbf{\hat{\eta}}-%
\mathbf{\eta }_{0})\overset{D}{\longrightarrow }N\left( \mathbf{0},\mathbf{%
\Omega }\right) \text{ as }n\rightarrow \infty ,
\end{equation*}%
where $\mathbf{\Omega }=\mathbf{W}^{-1}\mathbf{VW}^{-1}$, with $\mathbf{V}$
a $\left( dp+q\right) \times \left( dp+q\right) $ block-structured matrix 
\begin{equation}
\mathbf{V}=\left[ 
\begin{array}{cccc}
\mathbf{V}_{11} & \cdots & \mathbf{V}_{1d} & \mathbf{V}_{1,d+1} \\ 
\vdots & \ddots & \vdots & \vdots \\ 
\mathbf{V}_{d1} & \cdots & \mathbf{V}_{dd} & \mathbf{V}_{d,d+1} \\ 
\mathbf{V}_{d+1,1} & \cdots & \mathbf{V}_{d+1,d} & \mathbf{V}_{d+1,d+1}%
\end{array}%
\right]  \label{eq:V}
\end{equation}
with blocks 
\begin{equation*}
\mathbf{V}_{jj}=\frac{1}{d}\tilde{e}_{jj}\left( \mathbf{\Sigma }_{jj}-%
\mathbf{\sigma }_{j}\mathbf{\sigma }_{j}^{T}\right) +\frac{1}{d}\tilde{e}_{j}%
\mathbf{\Sigma }_{j},\text{ \ for }j=1,\ldots ,d,
\end{equation*}%
$\mathbf{V}_{jj^{\prime }}=\mathbf{O}$ for $j\neq j^{\prime }$, $j=1,\ldots
,d$, $j^{\prime }=1,\ldots ,d$, 
\begin{equation*}
\mathbf{V}_{j,d+1}=\frac{1}{d}\left( \mathbf{\sigma }_{jj}-\mathbf{\sigma }%
_{j}e_{j}\right) \mathbf{\tilde{\sigma}}_{jj}^{T}+\frac{1}{d}\mathbf{\sigma }%
_{j}\mathbf{\tilde{\sigma}}_{j}^{T},\text{ \ for }j=1,\ldots ,d,
\end{equation*}%
$\mathbf{V}_{d+1,j}=\mathbf{V}_{j,d+1}^{T}$ for $j=1,\ldots ,d$, and 
\begin{equation*}
\mathbf{V}_{d+1,d+1}=\frac{1}{d}\sum_{j=1}^{d}(e_{jj}-e_{j}^{2})\mathbf{%
\tilde{\Sigma}}_{jj}+\frac{1}{d}\sum_{j=1}^{d}e_{j}\mathbf{\tilde{\Sigma}}%
_{j}.
\end{equation*}%
Matrix $\mathbf{W}$ is also a $\left( dp+q\right) \times \left( dp+q\right) $
block-structured matrix similar to (\ref{eq:V}) with blocks 
\begin{equation*}
\mathbf{W}_{jj}=-\frac{1}{d}\tilde{e}_{j}\mathbf{\Sigma }_{j},\text{ \ for }%
j=1,\ldots ,d,
\end{equation*}%
$\mathbf{W}_{jj^{\prime }}=\mathbf{O}$ for $j\neq j^{\prime }$, $j=1,\ldots
,d$, $j^{\prime }=1,\ldots ,d$, 
\begin{equation*}
\mathbf{W}_{j,d+1}=-\frac{1}{d}\mathbf{\sigma }_{j}\mathbf{\tilde{\sigma}}%
_{j}^{T},\text{ \ for }j=1,\ldots ,d,
\end{equation*}%
$\mathbf{W}_{d+1,j}=\mathbf{W}_{j,d+1}^{T}$ for $j=1,\ldots ,d$, and 
\begin{equation*}
\mathbf{W}_{d+1,d+1}=-\frac{1}{d}\sum_{j=1}^{d}e_{j}\mathbf{\tilde{\Sigma}}%
_{j}.
\end{equation*}%
In the above expressions, 
\begin{equation*}
\tilde{e}_{j}=\frac{1}{w}\sum_{i=1}^{w}e^{\tilde{c}_{0}(a_{ij})},
\end{equation*}%
\begin{equation*}
\tilde{e}_{jj}=\frac{1}{w}\sum_{i=1}^{w}e^{2\tilde{c}_{0}(a_{ij})},
\end{equation*}%
\begin{equation*}
e_{jj}=\iint_{R\times R}\lambda _{0j}\left( u\right) \lambda _{0j}\left(
u^{\prime }\right) e^{\gamma _{0}\left( u,u^{\prime }\right) }du~du^{\prime
},
\end{equation*}%
\begin{equation*}
\mathbf{\sigma }_{jj}=\iint_{R\times R}\lambda _{0j}\left( u\right) \lambda
_{0j}\left( u^{\prime }\right) e^{\gamma _{0}\left( u,u^{\prime }\right) }%
\mathbf{\beta }\left( u\right) ~du~du^{\prime },
\end{equation*}%
\begin{equation*}
\mathbf{\tilde{\sigma}}_{j}=\frac{1}{w}\sum_{i=1}^{w}e^{\tilde{c}%
_{0}(a_{ij})}\mathbf{b}(a_{ij}),
\end{equation*}%
\begin{equation*}
\mathbf{\tilde{\sigma}}_{jj}=\frac{1}{w}\sum_{i=1}^{w}e^{2\tilde{c}%
_{0}(a_{ij})}\mathbf{b}(a_{ij}),
\end{equation*}%
\begin{equation*}
\mathbf{\Sigma }_{j}=\int_{R}\lambda _{0j}\left( u\right) \mathbf{\beta }%
\left( u\right) \mathbf{\beta }\left( u\right) ^{T}du,
\end{equation*}%
\begin{equation*}
\mathbf{\tilde{\Sigma}}_{j}=\frac{1}{w}\sum_{i=1}^{w}e^{\tilde{c}%
_{0}(a_{ij})}\mathbf{b}(a_{ij})\mathbf{b}(a_{ij})^{T},
\end{equation*}%
\begin{equation*}
\mathbf{\Sigma }_{jj}=\iint_{R\times R}\lambda _{0j}\left( u\right) \lambda
_{0j}\left( u^{\prime }\right) e^{\gamma _{0}\left( u,u^{\prime }\right) }%
\mathbf{\beta }\left( u\right) \mathbf{\beta }\left( u^{\prime }\right)
^{T}du~du^{\prime },
\end{equation*}%
\begin{equation*}
\mathbf{\tilde{\Sigma}}_{jj}=\frac{1}{w}\sum_{i=1}^{w}e^{2\tilde{c}%
_{0}(a_{ij})}\mathbf{b}(a_{ij})\mathbf{b}(a_{ij})^{T},
\end{equation*}%
\begin{equation*}
\mathbf{\Sigma }_{jj}=\iint_{R\times R}\lambda _{0j}\left( u\right) \lambda
_{0j}\left( u^{\prime }\right) e^{\gamma _{0}\left( u,u^{\prime }\right) }%
\mathbf{\beta }\left( u\right) \mathbf{\beta }\left( u^{\prime }\right)
^{T}du~du^{\prime },
\end{equation*}%
$\mathbf{\sigma }_{j}$ is as in (\ref{eq:sigma_j}) and $e_{j}$ is as in (\ref%
{eq:e_j}).
\end{theorem}

The asymptotic variance matrix $\mathbf{\Omega }$ in Theorem \ref%
{thm:AN_trend_plus_seas} follows the usual \textquotedblleft sandwich
formula\textquotedblright\ for M-estimators. When the true model coincides
with the working model, i.e.~when $Y_{t}\left( u\right) =0$, we have $\gamma
_{0}\left( u,u^{\prime }\right) =0$ and then $e_{jj}=e_{j}^{2}$, $\mathbf{%
\Sigma }_{jj}=\mathbf{\sigma }_{j}\mathbf{\sigma }_{j}^{T}$ and $\mathbf{%
\sigma }_{jj}=\mathbf{\sigma }_{j}e_{j}$, which implies $\mathbf{V}=-\mathbf{%
W}$ and then $\mathbf{\Omega }=\mathbf{V}^{-1}$, the inverse of Fisher's
information matrix for this model.

To estimate $\mathbf{\Omega }$ we need consistent estimators of $\mathbf{V}$
and $\mathbf{W}$. Plug-in estimators can be used for $\mathbf{W}$, because
the quantities $e_{j}$, $\tilde{e}_{j}$, $\mathbf{\sigma }_{j}$, $\mathbf{%
\tilde{\sigma}}_{j}$, $\mathbf{\Sigma }_{j}$ and $\mathbf{\tilde{\Sigma}}%
_{j} $ depend only on $\lambda _{j}\left( u\right) $ and $\tilde{c}%
_{j}\left( t\right) $, which can be consistently estimated by $\hat{\lambda}%
_{j}\left( u\right) $ and $\widehat{\tilde{c}}_{j}(t)$. But the quantities $%
\mathbf{\Sigma }_{jj}$, $\mathbf{\sigma }_{jj}$ and $e_{jj}$ that define $%
\mathbf{V}$ involve $\gamma _{0}\left( u,u^{\prime }\right) $, for which we
do not have a consistent estimator. However, it is shown in the proof of
Theorem \ref{thm:AN_trend_plus_seas} in the Supplementary Material that 
\begin{equation*}
\mathbf{V}=\lim\limits_{n\rightarrow \infty }\frac{1}{n}\sum_{t=1}^{n}E\{%
\mathbf{\psi }_{t}(X_{t},\mathbf{\theta }_{01}^{\ast },\ldots ,\mathbf{%
\theta }_{0d}^{\ast },\mathbf{\eta }_{0})\mathbf{\psi }_{t}(X_{t},\mathbf{%
\theta }_{01}^{\ast },\ldots ,\mathbf{\theta }_{0d}^{\ast },\mathbf{\eta }%
_{0})^{T}\},
\end{equation*}%
where $\mathbf{\psi }_{t}$ is the $\left( dp+q\right) $-dimensional
block-structured score function 
\begin{equation*}
\mathbf{\psi }_{t}\left( x_{t},\mathbf{\theta }_{1},\ldots ,\mathbf{\theta }%
_{d},\mathbf{\eta }\right) =\left[ 
\begin{array}{c}
\mathbf{\psi }_{t}\left( x_{t},\mathbf{\theta }_{1},\ldots ,\mathbf{\theta }%
_{d},\mathbf{\eta }\right) _{1} \\ 
\vdots \\ 
\mathbf{\psi }_{t}\left( x_{t},\mathbf{\theta }_{1},\ldots ,\mathbf{\theta }%
_{d},\mathbf{\eta }\right) _{d} \\ 
\mathbf{\psi }_{t}\left( x_{t},\mathbf{\theta }_{1},\ldots ,\mathbf{\theta }%
_{d},\mathbf{\eta }\right) _{d+1}%
\end{array}%
\right]
\end{equation*}%
with blocks 
\begin{equation*}
\mathbf{\psi }_{t}\left( x_{t},\mathbf{\theta }_{1},\ldots ,\mathbf{\theta }%
_{d},\mathbf{\eta }\right) _{j}=
\end{equation*}%
\begin{equation*}
-e^{\mathbf{\eta }^{T}\mathbf{b}(\{t\}_{r})}\delta _{j}\left( j\left(
t\right) \right) \int_{R}e^{\mathbf{\theta }_{j}^{T}\mathbf{\beta }\left(
u\right) }\mathbf{\beta }\left( u\right) du+\delta _{j}\left( j\left(
t\right) \right) \sum_{k=1}^{m_{t}}\mathbf{\beta }(u_{tk}),
\end{equation*}%
for $j=1,\ldots ,d$, and 
\begin{equation*}
\mathbf{\psi }_{t}\left( x_{t},\mathbf{\theta }_{1},\ldots ,\mathbf{\theta }%
_{d},\mathbf{\eta }\right) _{d+1}=
\end{equation*}%
\begin{equation*}
-e^{\mathbf{\eta }^{T}\mathbf{b}(\{t\}_{r})}\mathbf{b}(\{t\}_{r})\int_{R}e^{%
\mathbf{\theta }_{j(t)}^{T}\mathbf{\beta }\left( u\right) }du+m_{t}\mathbf{b}%
(\{t\}_{r}),
\end{equation*}%
where $\delta _{j}\left( \cdot \right) $ denotes Kronecker's delta function.
Then $\mathbf{V}$ can be consistently estimated by%
\begin{equation*}
\mathbf{\hat{V}}=\frac{1}{n}\sum_{t=1}^{n}\mathbf{\psi }_{t}(x_{t},\mathbf{%
\hat{\theta}}_{1},\ldots ,\mathbf{\hat{\theta}}_{d},\mathbf{\hat{\eta}})%
\mathbf{\psi }_{t}(x_{t},\mathbf{\hat{\theta}}_{1},\ldots ,\mathbf{\hat{%
\theta}}_{d},\mathbf{\hat{\eta}})^{T}.
\end{equation*}

Asymptotic confidence bands of level $1-\alpha $ for the trend $\tilde{c}%
_{0}\left( t\right) $ and the seasonal intensity functions $\lambda
_{0j}\left( u\right) $'s are given by 
\begin{equation*}
\widehat{\tilde{c}}(t)\pm z_{\alpha /2}\left\{ \frac{1}{n}\mathbf{b}(t)^{T}%
\mathbf{\hat{\Omega}}_{d+1,d+1}\mathbf{b}(t)\right\} ^{1/2}
\end{equation*}%
and 
\begin{equation*}
\hat{\lambda}_{j}\left( u\right) \pm z_{\alpha /2}\left\{ \frac{1}{n}\hat{%
\lambda}_{j}\left( u\right) \mathbf{\beta }\left( u\right) ^{T}\mathbf{\hat{%
\Omega}}_{jj}\mathbf{\beta }\left( u\right) \right\} ^{1/2},
\end{equation*}%
(the latter obtained after an application of the Delta Method) for $%
j=1,\ldots ,d$, where $\mathbf{\hat{\Omega}}_{d+1,d+1}$ and $\mathbf{\hat{%
\Omega}}_{jj}$ denote the respective blocks of $\mathbf{\hat{\Omega}}=%
\mathbf{\hat{W}}^{-1}\mathbf{\hat{V}\hat{W}}^{-1}$.

The asymptotics for the estimators $(\mathbf{\hat{\theta}},\mathbf{\hat{\eta}%
})$ of the trend-only model (\ref{eq:trend_model}) can be derived as
corollaries of Theorems \ref{thm:Consistency_trend_plus_seas} and \ref%
{thm:AN_trend_plus_seas}, adapting assumption A3 to specify that $\mu
_{0}\left( u\right) =\mathbf{\theta }_{0}^{T}\mathbf{\beta }\left( u\right) $
for $\mathbf{\theta }_{0}\in \mathbb{R}^{p}$ and $\lambda _{0}\left(
u\right) =\exp \left\{ \mu _{0}\left( u\right) +v_{0}\left( u\right)
/2\right\} $.

\begin{corollary}
\label{cor:Consistency_trend_only} Let $\mathbf{\theta }_{0}^{\ast }=\mathbf{%
\theta }_{0}+\mathbf{\tau }_{0}/2$ for $j=1,\ldots ,d$. Under assumptions
A1--A5 we have $\mathbf{\hat{\eta}}\overset{P}{\longrightarrow }\mathbf{\eta 
}_{0}$ and $\mathbf{\hat{\theta}}\overset{P}{\longrightarrow }\mathbf{\theta 
}_{0}^{\ast }$ as $n\rightarrow \infty $. Therefore $\widehat{\tilde{c}}(t)%
\overset{P}{\longrightarrow }\tilde{c}_{0}\left( t\right) $ and $\hat{\lambda%
}\left( u\right) \overset{P}{\longrightarrow }\lambda _{0}\left( u\right) $
pointwise as $n\rightarrow \infty $.
\end{corollary}

\begin{corollary}
\label{cor:AN_trend_only}Under assumptions A1--A6, we have 
\begin{equation*}
\sqrt{n}(\mathbf{\hat{\theta}}-\mathbf{\theta }_{0}^{\ast },\mathbf{\hat{\eta%
}}-\mathbf{\eta }_{0})\overset{D}{\longrightarrow }N\left( \mathbf{0},%
\mathbf{\Omega }\right) \text{ as }n\rightarrow \infty ,
\end{equation*}%
where $\mathbf{\Omega }=\mathbf{W}^{-1}\mathbf{VW}^{-1}$, with $\mathbf{V}$
a $\left( p+q\right) \times \left( p+q\right) $ block-structured matrix 
\begin{equation*}
\mathbf{V}=\left[ 
\begin{array}{cc}
\mathbf{V}_{11} & \mathbf{V}_{12} \\ 
\mathbf{V}_{21} & \mathbf{V}_{22}%
\end{array}%
\right]
\end{equation*}%
with blocks 
\begin{equation*}
\mathbf{V}_{11}=\tilde{e}_{00}\left( \mathbf{\Sigma }_{00}-\mathbf{\sigma }%
_{0}\mathbf{\sigma }_{0}^{T}\right) +\tilde{e}_{0}\mathbf{\Sigma }_{0},
\end{equation*}
\begin{equation*}
\mathbf{V}_{12}=\left( \mathbf{\sigma }_{00}-\mathbf{\sigma }%
_{0}e_{0}\right) \mathbf{\tilde{\sigma}}_{00}^{T}+\mathbf{\sigma }_{0}%
\mathbf{\tilde{\sigma}}_{0}^{T},
\end{equation*}%
$\mathbf{V}_{21}=\mathbf{V}_{12}^{T}$, and 
\begin{equation*}
\mathbf{V}_{22}=(e_{00}-e_{0}^{2})\mathbf{\tilde{\Sigma}}_{00}+e_{0}\mathbf{%
\tilde{\Sigma}}_{0}.
\end{equation*}%
Matrix $\mathbf{W}$ is also a $\left( p+q\right) \times \left( p+q\right) $
block-structured matrix with blocks 
\begin{equation*}
\mathbf{W}_{11}=-\tilde{e}_{0}\mathbf{\Sigma }_{0},
\end{equation*}
\begin{equation*}
\mathbf{W}_{12}=-\mathbf{\sigma }_{0}\mathbf{\tilde{\sigma}}_{0}^{T},
\end{equation*}%
$\mathbf{W}_{21}=\mathbf{W}_{12}^{T}$, and 
\begin{equation*}
\mathbf{W}_{22}=-e_{0}\mathbf{\tilde{\Sigma}}_{0}.
\end{equation*}%
In the above expressions, 
\begin{equation*}
\tilde{e}_{0}=\frac{1}{r}\sum_{i=1}^{r}e^{\tilde{c}_{0}(i)},
\end{equation*}%
\begin{equation*}
e_{0}=\int_{R}\lambda _{0}\left( u\right) du
\end{equation*}%
\begin{equation*}
\tilde{e}_{00}=\frac{1}{r}\sum_{i=1}^{r}e^{2\tilde{c}_{0}(i)},
\end{equation*}%
\begin{equation*}
e_{00}=\iint_{R\times R}\lambda _{0}\left( u\right) \lambda _{0}\left(
u^{\prime }\right) e^{\gamma _{0}\left( u,u^{\prime }\right) }du~du^{\prime
},
\end{equation*}%
\begin{equation*}
\mathbf{\sigma }_{00}=\iint_{R\times R}\lambda _{0}\left( u\right) \lambda
_{0}\left( u^{\prime }\right) e^{\gamma _{0}\left( u,u^{\prime }\right) }%
\mathbf{\beta }\left( u\right) ~du~du^{\prime },
\end{equation*}
\begin{equation*}
\mathbf{\sigma }_{0}=\int_{R}\lambda _{0}\left( u\right) \mathbf{\beta }(u)du
\end{equation*}%
\begin{equation*}
\mathbf{\tilde{\sigma}}_{0}=\frac{1}{r}\sum_{i=1}^{r}e^{\tilde{c}_{0}(i)}%
\mathbf{b}(i),
\end{equation*}%
\begin{equation*}
\mathbf{\tilde{\sigma}}_{00}=\frac{1}{r}\sum_{i=1}^{r}e^{2\tilde{c}_{0}(i)}%
\mathbf{b}(i),
\end{equation*}%
\begin{equation*}
\mathbf{\Sigma }_{0}=\int_{R}\lambda _{0}\left( u\right) \mathbf{\beta }%
\left( u\right) \mathbf{\beta }\left( u\right) ^{T}du,
\end{equation*}%
\begin{equation*}
\mathbf{\tilde{\Sigma}}_{0}=\frac{1}{r}\sum_{i=1}^{r}e^{\tilde{c}_{0}(i)}%
\mathbf{b}(i)\mathbf{b}(i)^{T},
\end{equation*}%
\begin{equation*}
\mathbf{\Sigma }_{00}=\iint_{R\times R}\lambda _{0}\left( u\right) \lambda
_{0}\left( u^{\prime }\right) e^{\gamma _{0}\left( u,u^{\prime }\right) }%
\mathbf{\beta }\left( u\right) \mathbf{\beta }\left( u^{\prime }\right)
^{T}du~du^{\prime },
\end{equation*}%
\begin{equation*}
\mathbf{\tilde{\Sigma}}_{00}=\frac{1}{r}\sum_{i=1}^{r}e^{2\tilde{c}_{0}(i)}%
\mathbf{b}(i)\mathbf{b}(i)^{T},
\end{equation*}%
and%
\begin{equation*}
\mathbf{\Sigma }_{00}=\iint_{R\times R}\lambda _{0}\left( u\right) \lambda
_{0}\left( u^{\prime }\right) e^{\gamma _{0}\left( u,u^{\prime }\right) }%
\mathbf{\beta }\left( u\right) \mathbf{\beta }\left( u^{\prime }\right)
^{T}du~du^{\prime }.
\end{equation*}
\end{corollary}

\section{Simulations\label{sec:Simulations}}

We ran some simulations to study the finite-sample behavior of the proposed
estimators. In order to generate realistic scenarios, the parameters for the
simulations were chosen to resemble the estimators obtained for the examples
in Section \ref{sec:Examples}.

We simulated the trend-only model (\ref{eq:trend_model}). For the $Y_{t}$s
we considered three situations: \emph{(i)} a model with $Y_{t}\left(
u\right) =0$, which is the working-likelihood model, \emph{(ii)} a model
with independent $Y_{t}$s, and \emph{(iii)} a model with autoregressive $%
Y_{t}$s.

For cases \emph{(ii)} and \emph{(iii)} we considered one-dimensional
Gaussian models of the form 
\begin{equation}
Y_{t}\left( u\right) =Z_{t}\zeta \left( u\right) ,  \label{eq:rank-one-model}
\end{equation}%
where $Z_{t}\sim N\left( 0,\sigma ^{2}\right) $ and $\zeta (u)$ is a
function of unit norm in $L^{2}\left( R\right) $. For case \emph{(ii)} the $%
Z_{t}$s were independent, and for case \emph{(iii)} they followed the
autoregressive model 
\begin{equation}
Z_{1}=\varepsilon _{1},\ \ Z_{t}=aZ_{t-1}+\varepsilon _{t}\text{ for }t\geq
2,  \label{eq:AR1_model}
\end{equation}%
where $\varepsilon _{t}\sim N\left( 0,\sigma _{\varepsilon }^{2}\right) $
and $\left\vert a\right\vert <1$; note that $\sigma ^{2}=\sigma
_{\varepsilon }^{2}/\left( 1-a^{2}\right) $ in this case.

To resemble the daily temporal processes analyzed in Section \ref%
{sec:Examples}, as range $R$ we took the interval $\left[ 0,24\right] $ and
as baseline mean $\mu _{0}\left( u\right) $ we took a cubic-spline function
with two equally-spaced knots and coefficients $\mathbf{\theta }_{0}=\left(
-5.45,-4.96,-0.13,-4.14,-1.15,-5.52\right) $. As time drift $c_{0}(t)$ we
took $c_{0}(t)=\tilde{c}_{0}\left( \frac{t}{n+1}\right) $ with $\tilde{c}%
_{0}\left( t\right) =9.38t-8.43t^{2}$ for $t\in \left[ 0,1\right] $. Note
that this parameterization for $c_{0}(t)$ is different from the one
mentioned in Section \ref{sec:Model}, but it is more convenient for making
comparisons across different sample sizes $n$. For model (\ref%
{eq:rank-one-model}) we took $\zeta \left( u\right) =\mu _{0}\left( u\right)
/\left\Vert \mu _{0}\right\Vert $, where $\left\Vert \cdot \right\Vert $
denotes the $L^{2}(R)$ norm; this situation, where the main direction of
variation is proportional to the mean, is common for functional data. To
generate daily counts with expected values similar to those of the examples
in Section \ref{sec:Examples}, which are around 13, we took $\sigma =2$ as
the standard deviation of the $Z_{t}$s for cases \emph{(ii)} and \emph{(iii)}%
. For the autoregressive model (\ref{eq:AR1_model}) we took $a=0.7$, which
implies $\sigma _{\varepsilon }=1.43$.

The variance function $v_{0}\left( u\right) =V\left\{ Y_{t}\left( u\right)
\right\} $ for model (\ref{eq:rank-one-model}) is $v_{0}\left( u\right)
=\sigma ^{2}\zeta ^{2}\left( u\right) $. Then, strictly speaking, $%
v_{0}\left( u\right) $ is not a cubic-spline function and assumption A4 in
Section \ref{sec:Asymptotics} is not strictly satisfied. However, the best
cubic-spline approximation to $v_{0}\left( u\right) $ has an $L^{2}(R)$
error of only $0.038$, so we will use the coefficients of this approximation
as $\mathbf{\tau }_{0}$, which gives $\mathbf{\tau }_{0}=\left(
0.508,0.331,-0.113,0.270,-0.056,0.459\right) $.

For each model \emph{(i)--(iii)} we considered three sample sizes: $n=50$
(small), $n=100$ (moderate) and $n=300$ (large). Each scenario was
replicated $1,000$ times. As measures of estimation error we computed bias,
standard deviation and root mean squared error of $\mathbf{\hat{\theta}}$, $%
\mathbf{\hat{\eta}}$, $\hat{\lambda}\left( u\right) $ and $\widehat{\tilde{c}%
}(t)$. For $\mathbf{\hat{\theta}}$, which converges to $\mathbf{\theta }%
_{0}^{\ast }=\mathbf{\theta }_{0}+\mathbf{\tau }_{0}/2$ by Corollary \ref%
{cor:Consistency_trend_only}, we defined $\mathrm{bias}(\mathbf{\hat{\theta})%
}=\Vert E(\mathbf{\hat{\theta}})-\mathbf{\theta }_{0}^{\ast }\Vert $, $%
\mathrm{sd}(\mathbf{\hat{\theta})}=\{E\Vert \mathbf{\hat{\theta}}-E(\mathbf{%
\hat{\theta}})\Vert ^{2}\}^{1/2}$ and $\mathrm{rmse}(\mathbf{\hat{\theta})}%
=\{E\Vert \mathbf{\hat{\theta}}-\mathbf{\theta }_{0}^{\ast }\Vert
^{2}\}^{1/2}$, where $\left\Vert \cdot \right\Vert $ is the Euclidean norm.
For $\mathbf{\hat{\eta}}$, $\hat{\lambda}\left( u\right) $ and $\widehat{%
\tilde{c}}(t)$ we used analogous definitions; for the functional estimators $%
\hat{\lambda}\left( u\right) $ and $\widehat{\tilde{c}}(t)$ the $L^{2}(R)$
and $L^{2}\left( \left[ 0,1\right] \right) $ norms were respectively used.

\begin{table}[tbp] \centering%

\begin{tabular}{llllccclccc}
&  &  &  & \multicolumn{3}{c}{Intensity} &  & \multicolumn{3}{c}{Trend} \\ 
\cline{5-7}\cline{9-11}
\multicolumn{1}{c}{Model} & \multicolumn{1}{c}{} & \multicolumn{1}{c}{$n$} & 
\multicolumn{1}{c}{} & bias & sd & rmse & \multicolumn{1}{c}{} & bias & sd & 
rmse \\ \hline
\multicolumn{1}{c}{(i)} & \multicolumn{1}{c}{} & \multicolumn{1}{c}{$50$} & 
\multicolumn{1}{c}{} & $.085$ & $.096$ & $.128$ & \multicolumn{1}{c}{} & $%
.129$ & $.173$ & $.216$ \\ 
\multicolumn{1}{c}{} & \multicolumn{1}{c}{} & \multicolumn{1}{c}{$100$} & 
\multicolumn{1}{c}{} & $.038$ & $.063$ & $.073$ & \multicolumn{1}{c}{} & $%
.059$ & $.125$ & $.138$ \\ 
\multicolumn{1}{c}{} & \multicolumn{1}{c}{} & \multicolumn{1}{c}{$300$} & 
\multicolumn{1}{c}{} & $.013$ & $.036$ & $.038$ & \multicolumn{1}{c}{} & $%
.020$ & $.076$ & $.078$ \\ 
\multicolumn{1}{c}{} & \multicolumn{1}{c}{} & \multicolumn{1}{c}{} & 
\multicolumn{1}{c}{} &  &  &  & \multicolumn{1}{c}{} &  &  &  \\ 
\multicolumn{1}{c}{(ii)} & \multicolumn{1}{c}{} & \multicolumn{1}{c}{$50$} & 
\multicolumn{1}{c}{} & $.090$ & $.133$ & $.160$ & \multicolumn{1}{c}{} & $%
.122$ & $.245$ & $.273$ \\ 
\multicolumn{1}{c}{} & \multicolumn{1}{c}{} & \multicolumn{1}{c}{$100$} & 
\multicolumn{1}{c}{} & $.041$ & $.087$ & $.097$ & \multicolumn{1}{c}{} & $%
.057$ & $.176$ & $.185$ \\ 
\multicolumn{1}{c}{} & \multicolumn{1}{c}{} & \multicolumn{1}{c}{$300$} & 
\multicolumn{1}{c}{} & $.016$ & $.047$ & $.049$ & \multicolumn{1}{c}{} & $%
.023$ & $.101$ & $.103$ \\ 
\multicolumn{1}{c}{} & \multicolumn{1}{c}{} & \multicolumn{1}{c}{} & 
\multicolumn{1}{c}{} &  &  &  & \multicolumn{1}{c}{} &  &  &  \\ 
\multicolumn{1}{c}{(iii)} & \multicolumn{1}{c}{} & \multicolumn{1}{c}{$50$}
& \multicolumn{1}{c}{} & $.093$ & $.206$ & $.226$ & \multicolumn{1}{c}{} & $%
.095$ & $.415$ & $.426$ \\ 
\multicolumn{1}{c}{} & \multicolumn{1}{c}{} & \multicolumn{1}{c}{$100$} & 
\multicolumn{1}{c}{} & $.046$ & $.148$ & $.155$ & \multicolumn{1}{c}{} & $%
.045$ & $.324$ & $.327$ \\ 
\multicolumn{1}{c}{} & \multicolumn{1}{c}{} & \multicolumn{1}{c}{$300$} & 
\multicolumn{1}{c}{} & $.014$ & $.081$ & $.083$ & \multicolumn{1}{c}{} & $%
.013$ & $.187$ & $.188$ \\ \hline
\end{tabular}

\caption{Simulation Results. Bias, standard deviation and root mean squared error for estimators of 
the baseline intensity function $\lambda(u)$ and the trend $\tilde{c}(t)$, for three different models: 
(i) $Y_t=0$, (ii) independent $Y_t$s, and (iii) autoregressive $Y_t$s.}\label%
{tab:simul_errors}%
\end{table}%

The results are shown in Table \ref{tab:simul_errors}. Only estimation
errors for $\hat{\lambda}\left( u\right) $ and $\widehat{\tilde{c}}(t)$ are
shown here; the results for $\mathbf{\hat{\theta}}$ and $\mathbf{\hat{\eta}}$
can be found in the Supplementary Material. Table \ref{tab:simul_errors}
shows that the biases decrease as $n$ increases and are comparable in
magnitude for the three models, confirming the consistency of the estimators
established in Corollary \ref{cor:Consistency_trend_only}. The standard
deviations of the estimators are comparatively larger than the biases and
tend to dominate the root mean squared error. The estimators have the
largest standard deviations under the autoregressive model \emph{(iii)}, as
expected, since the autocorrelations reduce the effective sample size of the
data compared to the independent model \emph{(ii)}. Nevertheless, it is
clear that, as $n$ increases, the standard deviations decrease, and the
estimators are consistent under the three models.

\section{An application: analysis of bike demand in bicycle-sharing systems 
\label{sec:Examples}}

Bicycle-sharing systems have become increasingly popular in large cities
around the world (Shaheen et al., 2010). These systems provide short-term
bicycle rental services at unattended stations distributed throughout the
city. For the system to run smoothly, it is necessary that both bicycles and
empty docks be available at every station. But a bike flow from a given area
of the city to another area is rarely matched by a simultaneous reverse
flow, which creates imbalances in the spatial distribution of the bikes
(Nair and Miller-Hooks, 2011). To manage this problem, bikes must be
manually relocated by trucks as part of the day-to-day operations of the
system. Therefore, understanding the spatiotemporal patterns of bike demand
is fundamental for the efficient planning and management of these systems.

As examples of temporal point-process time series, let us consider bicycle
check-out times at three stations in the Divvy bike-sharing system of
Chicago. The data is publicly available at the Chicago Data Portal,
https://data.cityofchicago.org. We will analyze trips that took place in
2016. For each bike station, $X_{t}$ is defined as the set of check-out
times for trips originating from that station on day $t$. We consider three
stations that are representative of different city areas: the station at the
intersection of Ashland Ave and Wrightwood Ave, in a residential area, the
station at the intersection of Lasalle St and Washington St in the Loop, the
city center, and the station at the Shedd Aquarium by the lake.

For each station, we fitted the trend-plus-seasonality model (\ref%
{eq:trend_plus_seas_model}) with seasonal period $d=7$, to reflect the
weekly effect that we expect to be present. For the trend $c\left( t\right) $
we used a cubic polynomial, and for the seasonal means $\mu _{j}\left(
u\right) $ we used cubic splines with four equally-spaced knots on $\left(
0,24\right) $. Figures \ref{fig:plot_AW}--\ref{fig:plot_Shedd} show the
exponentiated trends $\exp \hat{c}\left( t\right) $ and the seasonal
intensities $\hat{\lambda}_{j}\left( u\right) $'s for the three stations.
The weekday seasonal intensities are plotted as solid lines and the weekend
intensities as dashed lines.

\FRAME{ftbpFU}{5.6299in}{3.2197in}{0pt}{\Qcb{Divvy data analysis, Ashland \&
Wrightwood station. (a) Exponentiated trend, (b) seasonal intensities
(solid: weekdays; dashed: weekends).}}{\Qlb{fig:plot_AW}}{plot_aw.eps}{%
\special{language "Scientific Word";type "GRAPHIC";maintain-aspect-ratio
TRUE;display "ICON";valid_file "F";width 5.6299in;height 3.2197in;depth
0pt;original-width 8.003in;original-height 5.9949in;cropleft "0";croptop
"0.8802";cropright "1";cropbottom "0.1195";filename
'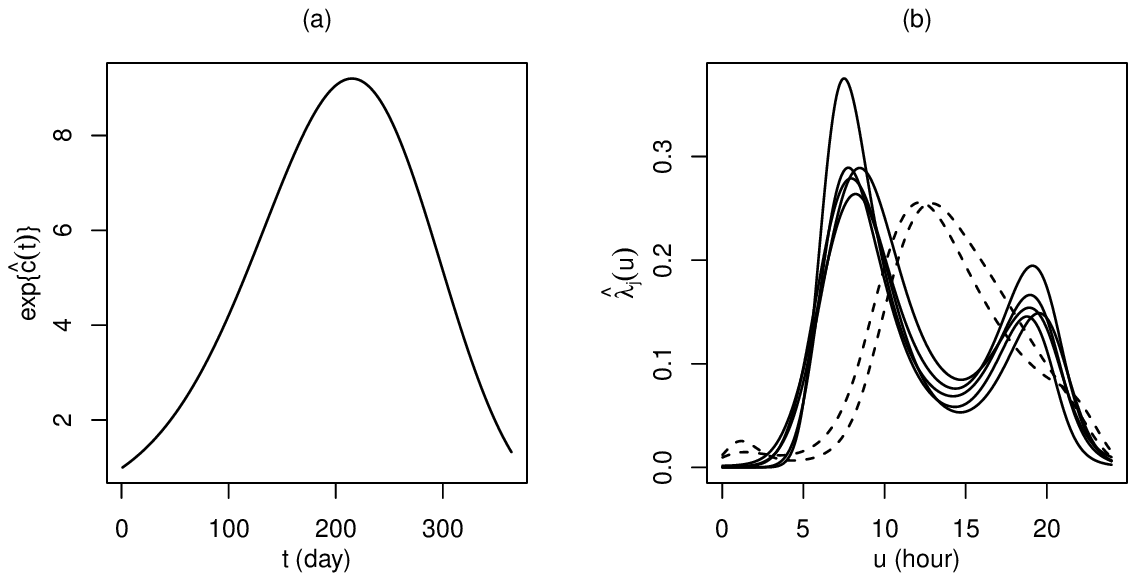';file-properties "XNPEU";}}

\FRAME{ftbpFU}{5.6299in}{3.2197in}{0pt}{\Qcb{Divvy data analysis, Lasalle \&
Washington station. (a) Exponentiated trend, (b) seasonal intensities
(solid: weekdays; dashed: weekends).}}{\Qlb{fig:plot_Las}}{plot_las.eps}{%
\special{language "Scientific Word";type "GRAPHIC";maintain-aspect-ratio
TRUE;display "ICON";valid_file "F";width 5.6299in;height 3.2197in;depth
0pt;original-width 8.003in;original-height 5.9949in;cropleft "0";croptop
"0.8803";cropright "1";cropbottom "0.1196";filename
'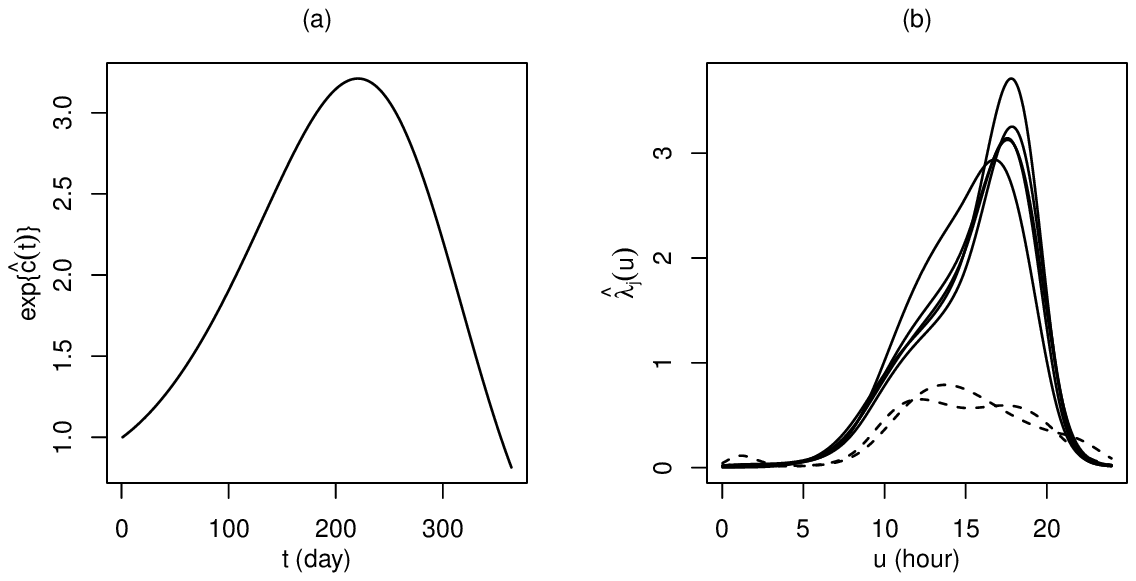';file-properties "XNPEU";}}

\FRAME{ftbpFU}{5.6299in}{3.2197in}{0pt}{\Qcb{Divvy data analysis, Shedd
Acquarium station. (a) Exponentiated trend, (b) seasonal intensities (solid:
weekdays; dashed: weekends).}}{\Qlb{fig:plot_Shedd}}{plot_shedd.eps}{\special%
{language "Scientific Word";type "GRAPHIC";maintain-aspect-ratio
TRUE;display "ICON";valid_file "F";width 5.6299in;height 3.2197in;depth
0pt;original-width 8.003in;original-height 5.9949in;cropleft "0";croptop
"0.8803";cropright "1";cropbottom "0.1196";filename
'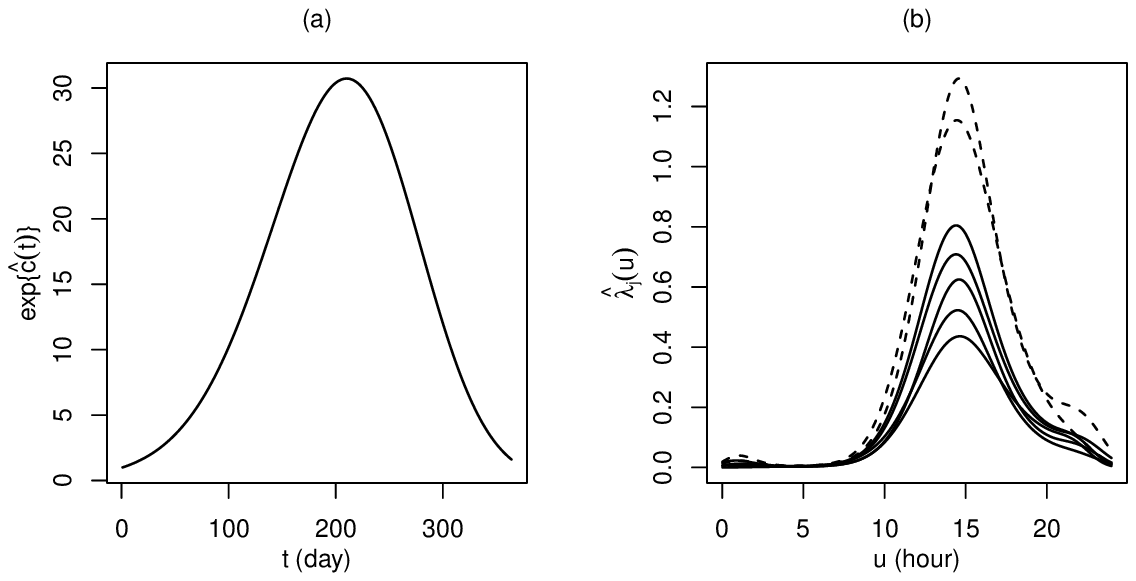';file-properties "XNPEU";}}

The shape of the exponentiated trend $\exp \hat{c}\left( t\right) $ is
similar for the three stations, as expected, since it essentially follows
the annual temperature cycle. But the magnitudes are very different. The
bike demand at the Shedd Aquarium station experiences a thirty-fold increase
in summer compared to winter (Fig.~\ref{fig:plot_Shedd}(a)), whereas the
residential Ashland \& Wrightwood station experiences a nine-fold increase
(Fig.\ \ref{fig:plot_AW}(a)), and the downtown Lasalle \& Washington station
only a three-fold increase (Fig.~\ref{fig:plot_Las}(a)).

The shapes of the seasonal baseline intensities vary widely from station to
station and between weekdays and weekends, reflecting three characteristic
but very different patterns of bike demand. The residential pattern,
represented by the Ashland \& Wrightwood station (Fig.\ \ref{fig:plot_AW}%
(b)), shows peaks at 8am and 6pm on weekdays, corresponding to the morning
and evening commutes, and a single peak at noon on weekends. The downtown
pattern, represented by the Lasalle \& Washington station (Fig.~\ref%
{fig:plot_Las}(b)), shows a single peak at 6pm on weekdays, corresponding to
the evening commute, and a flat and relatively low bike demand on weekends.
The touristic pattern, represented by the Shedd Aquarium station (Fig.~\ref%
{fig:plot_Shedd}(b)), shows a single peak at 3pm for all days of the week,
but the peak is higher for weekends than for weekdays.

\section*{Acknowledgement}

This work was partly supported by NSF grant DMS 2412015.

\section*{References}

\begin{description}
\item Baddeley, A. (2007). Spatial point processes and their applications.
In \emph{Stochastic Geometry}, \emph{Lecture Notes in Mathematics} \textbf{%
1892} 1--75. Springer, New York.

\item Begu, B., Panzeri, S., Arnone, E., Carey, M., and Sangalli, L.M.
(2024). A nonparametric penalized likelihood approach to density estimation
of space--time point patterns. \emph{Spatial Statistics} \textbf{61 }100824.

\item Bouzas, P.R., Valderrama, M., Aguilera, A.M., and Ruiz-Fuentes, N.
(2006). Modelling the mean of a doubly stochastic Poisson process by
functional data analysis. \emph{Computational Statistics and Data Analysis} 
\textbf{50} 2655--2667.

\item Bouzas, P.R., Ruiz-Fuentes, N., and Oca\~{n}a, F.M. (2007). Functional
approach to the random mean of a compound Cox process. \emph{Computational
Statistics }\textbf{22} 467--479.

\item Brockwell, P.J., and Davis, R.A. (2006). \emph{Time Series: Theory and
Methods (2nd ed.).} Springer, New York.

\item Cox, D.R., and Isham, V. (1980). \emph{Point Processes.} Chapman and
Hall/CRC, Boca Raton.

\item Diggle, P.J. (2013). \emph{Statistical Analysis of Spatial and
Spatio-Temporal Point Patterns, Third Edition.} Chapman and Hall/CRC, Boca
Raton.

\item Diggle, P., Rowlingson, B. and Su, T.-l. (2005). Point process
methodology for on-line spatio-temporal disease surveillance. \emph{%
Environmetrics} \textbf{16} 423--434.

\item Diggle, P. J., Moraga, P., Rowlingson, B., and Taylor, B. M. (2013).
Spatial and spatio-temporal log-Gaussian Cox processes: extending the
geostatistical paradigm. \emph{Statistical Science} \textbf{28} 542--563.

\item Fern\'{a}ndez-Alcal\'{a}, R.M., Navarro-Moreno, J., and Ruiz-Molina,
J.C. (2012). On the estimation problem for the intensity of a DSMPP. \emph{%
Methodology and Computing in Applied Probability }\textbf{14} 5--16.

\item Gajardo, \'{A}., and M\"{u}ller, H. G. (2022). Cox point process
regression. \emph{IEEE Transactions on Information Theory} \textbf{68}
1133--1156.

\item Gajardo, \'{A}., and M\"{u}ller, H. G. (2023). Point process models
for COVID-19 cases and deaths. \emph{Journal of Applied Statistics} \textbf{%
50} 2294--2309.

\item Gajardo, \'{A}, M\"{u}ller, H. G., and Zhou, H. (2025). Wasserstein-Fr%
\'{e}chet integration of conditional distributions. \emph{Electronic Journal
of Statistics} \textbf{19} 1722--1783.

\item Gervini, D. (2016). Independent component models for replicated point
processes. \emph{Spatial Statistics} \textbf{18} 474--488.

\item Gervini, D. (2022a). Doubly stochastic models for spatio-temporal
covariation of replicated point processes. \emph{Canadian Journal of
Statistics} \textbf{50} 287--303.

\item Gervini, D. (2022b). Spatial kriging for replicated temporal point
processes. \emph{Spatial Statistics }\textbf{51} 100681.

\item Gervini, D. and Khanal, M. (2019). Exploring patterns of demand in
bike sharing systems via replicated point process models. \emph{Journal of
the Royal Statistical Society Series C: Applied Statistics} \textbf{68}
585--602.

\item Gervini, D. and Baur, T.J. (2020). Joint models for grid point and
response processes in longitudinal and functional data. \emph{Statistica
Sinica} \textbf{30} 1905--1924.

\item Gervini, D. (2025). Autocorrelation functions for point-process time
series. \emph{Journal of Time Series Analysis}.
https://doi.org/10.1111/jtsa.70018.

\item Horv\'{a}th, L., and Kokoszka, P. (2012). \emph{Inference for
Functional Data with Applications.} Springer, New York.

\item Huang, K., Chen, X., Guan, Y., and Li, Y. (2025). Partially linear
single-index models and functional principal component analysis of spatially
and temporally indexed point processes. \emph{Journal of Business and
Economic Statistics} \textbf{43} 1077--1091.

\item Huber, P.J., and Ronchetti, E.M. (2009). \emph{Robust Statistics (2nd
ed.)} John Wiley \& Sons, Hoboken, NJ.

\item Marron, J.S., and Dryden, I.L. (2021). \emph{Object Oriented Data
Analysis (1st ed.)}. Chapman and Hall/CRC.

\item M\o ller, J., Syversveen, A.R., and Waagepetersen, R.P. (1998). Log
Gaussian Cox processes. \emph{Scandinavian Journal of Statistics} \textbf{25}
451--482.

\item M\o ller, J., and Waagepetersen, R.P. (2004). \emph{Statistical
Inference and Simulation for Spatial Point Processes}. Chapman and Hall/CRC,
Boca Raton.

\item Nair, R., and Miller-Hooks, E. (2011). Fleet management for vehicle
sharing operations. \emph{Transportation Science }\textbf{45 }524--540.

\item Pollard, D. (1984). \emph{Convergence of Stochastic Processes.}
Springer, New York.

\item Qiu, J., Dai, X., \& Zhu, Z. (2024). Nonparametric estimation of
repeated densities with heterogeneous sample sizes. \emph{Journal of the
American Statistical Association} \textbf{119} 176--188.

\item Ramsay, J.O., and Silverman, B.W. (2005).\emph{\ Functional Data
Analysis (2nd ed.). }Springer, New York.

\item Ratcliffe, J. (2010). Crime mapping: spatial and temporal challenges.
In \emph{Handbook of Quantitative Criminology}, A.R. Piquero and D. Weisburd
(eds.), pp.~5--24. Springer, New York.

\item Shaheen, S., Guzman, S., and Zhang, H. (2010). Bike sharing in Europe,
the Americas and Asia: Past, present and future. \emph{Transportation
Research Record: Journal of the Transportation Research Board }\textbf{2143}
159--167.

\item Snyder, D.L., and Miller, M.I. (1991). \emph{Random Point Processes in
Time and Space.} Springer, New York.

\item Streit, R.L. (2010). \emph{Poisson Point Processes: Imaging, Tracking,
and Sensing.} Springer, New York.

\item Wu, S., M\"{u}ller, H.-G., and Zhang, Z. (2013). Functional data
analysis for point processes with rare events. \emph{Statistica Sinica }%
\textbf{23} 1--23.

\item Xu, G., Wang, M., Bian, J., Huang, H., Burch, T. R., Andrade, S. C.,
Zhang, J., and Guan, Y. (2020a). Semi-parametric learning of structured
temporal point processes. \emph{Journal of Machine Learning Research }%
\textbf{21} 1--39.

\item Xu, G., Zhao, C., Jalilian, A., Waagepetersen, R., Zhang, J., and
Guan, Y. (2020b). Nonparametric estimation of the pair correlation function
of replicated inhomogeneous point processes. \emph{Electronic Journal of
Statistics} \textbf{14} 3730--3765.
\end{description}

\end{document}